\documentclass[twocolumn]{aastex62}

\newcommand\msun{M$_{\odot}$}
\newcommand{\sci}[2]{#1 $\times$ 10$^{#2}$}

\accepted{August 9, 2019}

\submitjournal{The Astrophysical Journal Letters}

\shorttitle{Resolved circumbinary ring around L1551 IRS 5}
\shortauthors{Cruz-S\'aenz de Miera et al.}

\begin{document}

\title{Resolved ALMA continuum image of the circumbinary ring and circumstellar disks in the L1551 IRS 5 system}

\correspondingauthor{Fernando Cruz-S\'aenz de Miera}
\email{cruzsaenz.fernando@csfk.mta.hu}

\author{Fernando Cruz-S\'aenz de Miera}
\affil{Konkoly Observatory, Research Centre for Astronomy and Earth Sciences,\\ Hungarian Academy of Sciences, Konkoly-Thege Mikl\'os \'ut 15-17, 1121 Budapest, Hungary}

\author{\'Agnes K\'osp\'al}
\affil{Konkoly Observatory, Research Centre for Astronomy and Earth Sciences,\\ Hungarian Academy of Sciences, Konkoly-Thege Mikl\'os \'ut 15-17, 1121 Budapest, Hungary}\affiliation{Max Planck Institute for Astronomy, K\"onigstuhl 17,
  69117 Heidelberg, Germany}

\author{P\'eter \'Abrah\'am}
\affil{Konkoly Observatory, Research Centre for Astronomy and Earth Sciences,\\ Hungarian Academy of Sciences, Konkoly-Thege Mikl\'os \'ut 15-17, 1121 Budapest, Hungary}

\author{Hauyu Baobab Liu}
\affil{Institute of Astronomy and Astrophysics, Academia Sinica, 11F of Astronomy-Mathematics Building,\\ AS/NTU, No.1, Sec. 4, Roosevelt Rd., Taipei 10617, Taiwan, R.O.C.}

\author{Michihiro Takami}
\affil{Institute of Astronomy and Astrophysics, Academia Sinica, 11F of Astronomy-Mathematics Building,\\ AS/NTU, No.1, Sec. 4, Roosevelt Rd., Taipei 10617, Taiwan, R.O.C.}

\begin{abstract}
L1551 IRS 5 is a FUor-like object located in the Taurus star forming region.
We present ALMA 1.3 mm continuum observations using a wide range of baselines.
The observations recovered the two circumstellar disks composing the system and, for the first time, resolved the circumbinary ring.
We determined the geometry and estimated lower mass limits for the circumstellar disks using simple models.
We calculated lower limits for the total mass of both circumstellar disks.
After subtracting the two circumstellar disk models from the image, the residuals show a clearly resolved circumbinary ring.
Using a radiative transfer model, we show that geometrical effects can explain some of the brightness asymmetries found in the ring.
The remaining features are interpreted as enhancements in the dust density.
\end{abstract}

\keywords{stars: pre-main sequence --- circumstellar matter --- stars: individual(L1551 IRS 5)}

\section{Introduction}\label{sec:intro}
Stellar multiplicity is widespread: although the multiplicity fraction decreases with age, it is possible that all stars are born in multiple systems \citep{Duchene2013,Reipurth2014}.
Some of these break up later due to dynamical interactions, but some survive for long enough to form circumbinary planets \citep{welsh2015}.
Such planets form in protoplanetary disks which may be substantially affected by the components of the stellar multiple system.
For example, depending on the separation, disks may be truncated from the outside (in case of a circumstellar disk) or from the inside (in case of a circumbinary disk, \citealt{Artymowicz1994}).
Eccentric binaries may induce spiral arms in the circumbinary disk \citep{Mayama2010,Takakuwa2017ApJ...837...86T,Wagner2018}.
While disk truncation may inhibit planet formation, there are indications that circumbinary disks may feed material to circumstellar disks, helping planet formation \citep{dutrey2014,boehler2017}.
All these processes and structures may affect grain growth and, consequently, planet formation \citep{Price2018}.

In order to understand the details of disk evolution and planet formation in binary and multiple systems, observations of the disks' thermal emission are indispensable.
If these observations have high spatial resolution, they can reveal the complex morphology of the circumstellar and circumbinary material, while data at multiple wavelengths help constrain the optical depth and the grain size distribution \citep{Birnstiel2010,Testi2014,Woitke2016}.
Here we present \mbox{1.3 mm}, high spatial resolution observations of the young binary L1551~IRS~5 taken with the Atacama Large Millimeter/submillimeter Array (ALMA).
We analyze the structure of the cold material in the system at \mbox{20 au} resolution.

\section{Our target} \label{sec:target}
L1551 IRS 5 is a deeply embedded very young \mbox{($\sim$10$^5$ yr)} system at a distance of 147 $\pm$ 5 pc classified as a FUor-like object by \cite{Connelley2018ApJ...861..145C}.
It consists of a 0.8 \msun{} primary (N component), which dominates the system's luminosity, and a 0.3 \msun{} secondary (S component), which is invisible at near-infrared wavelengths \citep{Duchene2007}.
The separation of the binary is 0$\farcs$36 or \mbox{50 au} \citep{Liseau2005,Lim2016ApJ...826..153L}.
Its total bolometric luminosity is in the range of 25 -- 40 $L_{\odot}$ \citep{Osorio2003ApJ...586.1148O, Liseau2005}.
Both stars accrete at a much higher rate than normal T Tauri stars \citep[$\sim$10$^{-8}$ \msun{} yr$^{-1}$; e.g.][]{Gullbring1998ApJ...492..323G} with accretion rates of \mbox{$\dot{M_N}$ = \sci{6}{-6} \msun{} yr$^{-1}$} and \mbox{$\dot{M_S}$ = \sci{2}{-6} \msun{} yr$^{-1}$} \citep{Liseau2005}.
The sources also drive powerful radio jets observed at \mbox{7 mm} \citep{Lim2016ApJ...826..153L} and \mbox{3.5 cm} \citep{Rodriguez2003ApJ...586L.137R}.
According to the modeling of \cite{Osorio2003ApJ...586.1148O} the stars are surrounded by small, but massive circumstellar disks (CS disk masses: M$_{\rm disk,N}=0.25$ \msun{}, M$_{\rm disk,S}=0.1$ \msun{}, CS disk radii about \mbox{12 au} at \mbox{7 mm}), a circumbinary ring (CB ring mass: 0.4 \msun{}), and an envelope (envelope mass: 4 \msun{}).
Both CS disks have similar inclinations (45\degr) and position angles (150\degr, \citealt{Lim2016ApJ...826..153L}).

\section{Observations and data reduction} \label{sec:observations}
We observed L1551 IRS 5 in ALMA Cycle 4 (ID: 2016.1.00209.S; PI: Takami) with the 7\,m array using 11 antennas with projected baselines between 8.9 and 45 m on 2016 November 14, with the 12\,m array in compact configuration using 44 antennas with projected baselines between 15 and 500 m on 2018 April 24, and in extended configuration using  44 antennas with projected baselines between 30.5 m and 3.6 km on 2017 July 24. Data were taken in Band 6, with three spectral windows centered at the $^{12}$CO (at 30 kHz resolution), $^{13}$CO and C$^{18}$O $J=2-1$ lines (at 60\,kHz resolution), and  two continuum spectral windows of 1875 MHz bandwidth each, centered at 216.9\,GHz and 232.2\,GHz. Here we present the results of the continuum data; the analysis of line data will be presented in a later publication. The data sets were calibrated using CASA v.5.1.1 \citep{McMullin2007ASPC..376..127M}. We applied self-calibration for both the visibility phases and  amplitudes to improve signal-to-noise ratio. The common baselines between the three different observing configurations were consistent, therefore we concatenated them into a single data set which  recovers the emission in all spatial scales.
After excluding the channels with line emission from the combined data set, we created continuum images by cleaning using uniform and natural weighting.
For the uniform image the resulting synthesized beam size is \mbox{0$\farcs$144 $\times$ 0$\farcs$102} \mbox{(21 au $\times$ 15 au)} while the rms noise is \mbox{0.33 mJy beam$^{-1}$}.
In the case of the natural image the resulting synthesized beam size is \mbox{0$\farcs$198 $\times$ 0$\farcs$183} \mbox{(29 au $\times$ 27 au)} while the rms noise is \mbox{0.76 mJy beam$^{-1}$}.

\section{Structures in the continuum map} \label{sec:results}
\begin{figure*}[ht]
\plotone{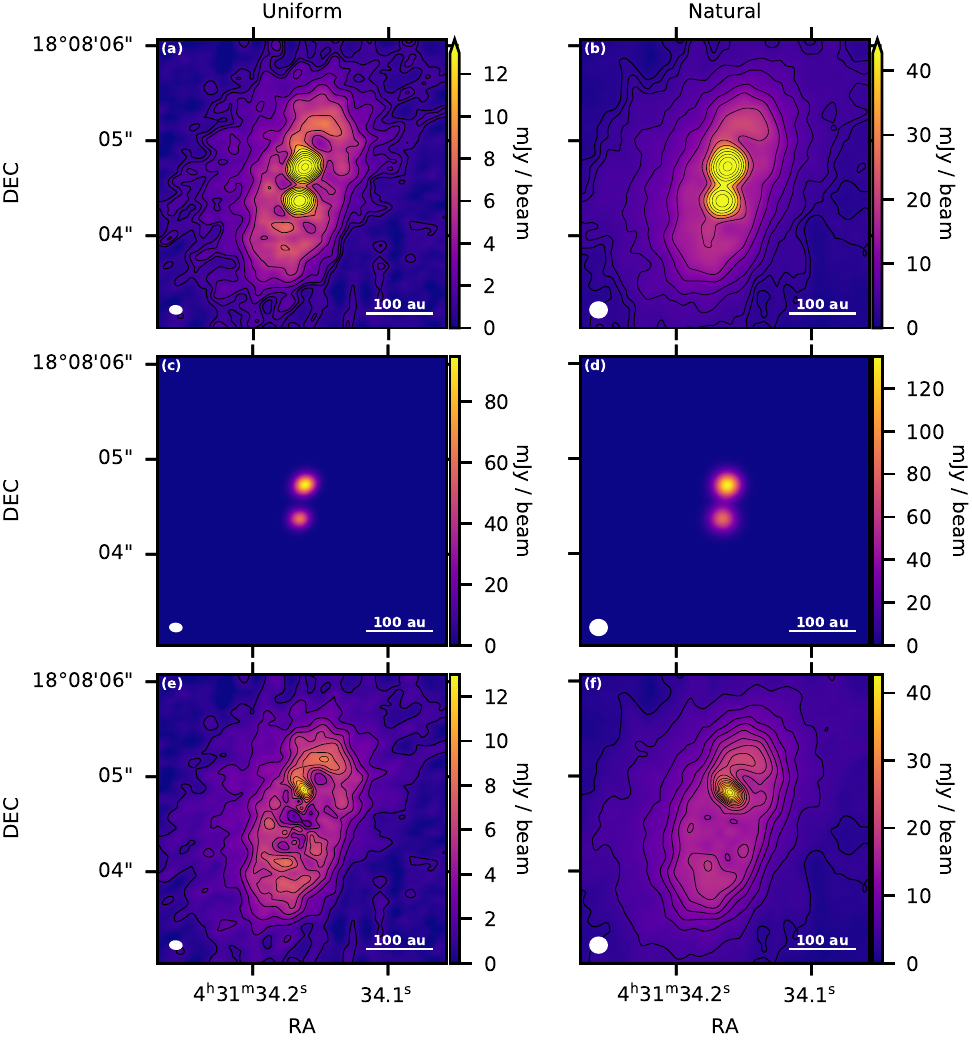}
\caption{1.3 mm continuum ALMA images of L1551 IRS 5.
The left and right columns are the uniform and natural weighted images, respectively.
The synthesized beam is shown in the bottom left corner of each panel.
Panels (a, b) show the complete flux range in logarithmic scale.
The uniform contours range from 3 to 283$\sigma$ and the natural contours range from 3 to 205$\sigma$, in both cases the contours are in logarithmic steps.
Panels (c, d) show the 2D Gaussian disk models for the CS disks.
Panels (e, f) show the residuals after subtracting the contribution from the CS disks.
The uniform contours range from 3 to 36$\sigma$ and the natural contours range from 3 to 56$\sigma$, in both cases the contours are in steps of 3$\sigma$.
\label{fig:map}}
\end{figure*}

\autoref{fig:map} (a, b) show the 1.3 mm continuum emission map of L1551 IRS 5 with the uniform and natural weighting schemes, respectively.
The binary system and its circumbinary material are clearly detected.
Based on the high SNR of our detections, we aimed to estimate geometrical properties of the CS disks based on the deconvolution of 2D Gaussians.
We used CASA's IMFIT tool to fit both CS disks simultaneously using a single 2D Gaussian for each disk.
We estimated the disks radii from the half-width half-maximum assuming a distance to the system of 147 $\pm$ 5 pc.
The disk inclinations were calculated from the ratio between the semi-axes and the position angles were directly taken from the deconvolved parameters.
Our results for the two CS disks are presented in \autoref{tab:gauss}.
We found that both disks are marginally resolved and their estimated radii indicate that both CS disks are fairly compact.
The resulting inclination and position angles are better constrained in the N disk, which is not surprising due to its slightly larger radius \citep{Lim2016ApJ...826..153L}.
However, as we show below, there are indications that the emission from our CS disks is mixed with other components, which cannot be resolved with the angular resolution of our current observations, and cause the large uncertainties in our fitting results.
From the centroids of our two-dimensional Gaussian function we estimated the separation between the two protostars to be 365 $\pm$ 4 mas \mbox{(51 $\pm$ 0.6 au)}; this separation is in agreement with the orbital analysis by \citet{Lim2016ApJ...826..153L}.

After removing the contribution from the fitted 2D Gaussians, the residuals showed that a single component was not enough to describe the emission from each CS disk, similar to the case of the 7 mm VLA observations of L1551 IRS 5 \citep{Lim2016ApJ...826..153L}.
We followed a procedure similar to \citet{Lim2016ApJ...826..153L} and used two 2D Gaussians to describe each circumstellar disk and found that the positions of the components are similar to those found in \citet{Lim2016ApJ...826..153L}, which were attributed to jet emission.
However, due to the beam size of our observations we are not able to clearly disentangle the emission from the CS disks and possible contribution from the jets; therefore, we cannot make a proper comparison between our results and those of \citet{Lim2016ApJ...826..153L}.

In Panels (e) and (f) of \autoref{fig:map} we show the residuals after subtracting the four best-fitted 2D Gaussians describing the CS disks.
In both residuals, we see extended circumbinary emission which had been reported in previous works \citep[e.g.][]{Chou2014ApJ...796...70C, Liu2018A&A...612A..54L}.
However, the present ones are the first resolved observations of the circumbinary material.
We estimated the radius, inclination and position angle of the extended emission by fitting an ellipse to all pixels above 12$\sigma$ in the residual images.
These results are also shown in \autoref{tab:gauss}.

Residuals in both the uniformly and naturally weighted maps show a bright source slightly to the north of the position of the CS N disk.
It has a peak emission of $\sim$13 mJy beam$^{-1}$ and $\sim$43 mJy beam$^{-1}$ for the uniform and natural residuals, respectively.
This source had not been detected previously, further analysis of this component is beyond the scope of this letter and will be presented in a follow-up paper.

The residuals show less emission in the center of the image, indicating the presence on a circumbinary ring.
This is more clearly seen in the uniform residuals (e), than in the natural residuals (f), due to the difference in beam sizes.
However, in both cases, the northern region of the CB ring is brighter than the southern region, which appears to be more extended.

\begin{figure*}[ht]
\plotone{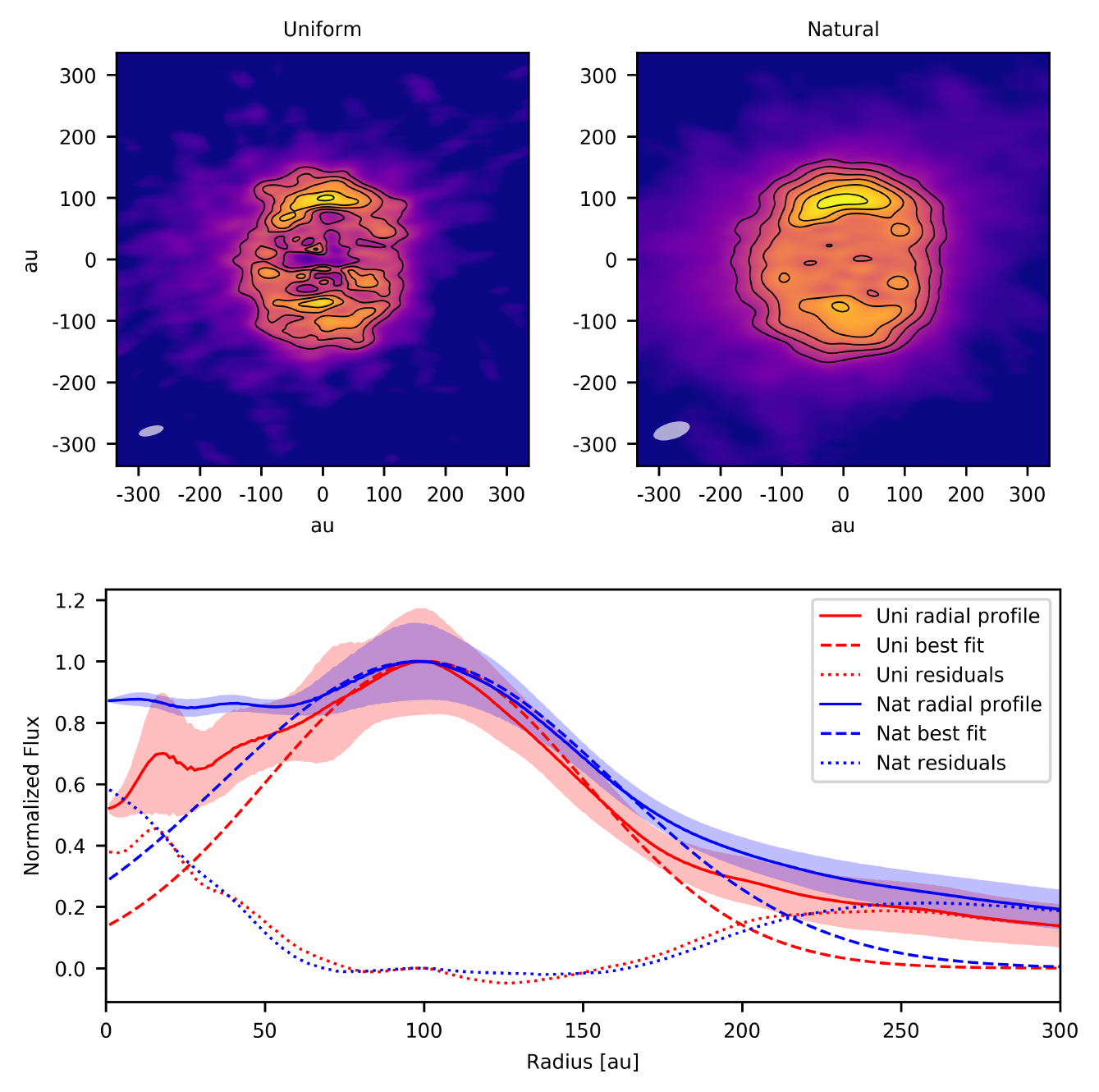}
\caption{\emph{Top row:} Deprojected maps of the uniform and natural residuals, after the removal of the CS disks and the remaining bright source. The contours range from 12$\sigma$ to 24$\sigma$ and 27$\sigma$ for the uniform and natural images, respectively, with steps of 3$\sigma$. \emph{Bottom row:} Radial profiles of the deprojected residuals (solid lines), best fitted 1D Gaussian (dashed lines) and the residuals (dotted lines).
\label{fig:deprojected_rp}}
\end{figure*}

In the top row of \autoref{fig:deprojected_rp} we show the deprojected residuals for both weighting schemes.
We began by using a 2D Gaussian to remove the bright source from panels (e) and (f) of \autoref{fig:map}, we rotated the remaining emission based on the position angle of the circumbinary emission and deprojected the residuals using the inclination angles of the CB rings.
In the bottom row of \autoref{fig:deprojected_rp} we show the azimuthally averaged radial profiles of both deprojected residuals.
We estimated the width of the ring by fitting a 1D Gaussian to the peak of the radial profiles and deconvolved them by their respective beams.
We found that the ring peaks at 100 $\pm$ 1 au and has a width of 38 $\pm$ 0.6 au.

In both weighting schemes, the residuals show the CB ring surrounded by faint emission which is interpreted as the dust contribution from the system's extended envelope.

\begin{deluxetable*}{ccccccc}[ht]
\tablecaption{Modeling results of the components of the L1551 IRS 5 system.\label{tab:gauss}}
\tablecolumns{8}
\tablewidth{0pt}
\tablehead{
\colhead{Weighting}&
\colhead{Radius}&
\colhead{Inclination}&
\colhead{Position angle}&
\colhead{Peak flux}&
\colhead{Integrated flux}&
\colhead{Disk mass}\\
\colhead{}&
\colhead{(au)}&
\colhead{(deg)}&
\colhead{(deg)}&
\colhead{(mJy beam$^{-1}$)}&
\colhead{(mJy)}&
\colhead{(\msun)}
}
\startdata
\multicolumn{7}{c}{North}\\
\tableline
Natural & 13.3 $\pm$ 3.5 & 35 $\pm$ 14 & 169 $\pm$ 31 & 146.0 $\pm$ 7.0 & 304 $\pm$ 22 & $>$\sci{6.03}{-3}\\
Uniform & 10.3 $\pm$ 1.8 & 40 $\pm$ 9 & 156 $\pm$ 16 & 92.5 $\pm$ 2.9 & 236 $\pm$ 10 & $>$\sci{4.68}{-3}\\
\tableline
\multicolumn{7}{c}{South}\\
\tableline
Natural & 14.0 $\pm$ 5.0 & 45 $\pm$ 15 & 80 $\pm$ 60 & 89.0 $\pm$ 7.0 & 178 $\pm$ 21 & $>$\sci{5.76}{-3}\\
Uniform &  8.8 $\pm$ 2.7 & 24 $\pm$ 25 & 110 $\pm$ 70 & 59.7 $\pm$ 2.9 & 137 $\pm$ 9 & $>$\sci{4.44}{-3}\\
\tableline
\multicolumn{7}{c}{Circumbinary Ring}\\
\tableline
Natural & 141.9 $\pm$ 7.4 & 61.5 $\pm$ 1.7 & 162 $\pm$ 2 & $-$ & 729.9 $\pm$ 81.9 & \sci{2.99}{-2}\\
Uniform & 136.8 $\pm$ 6.1 & 59.5 $\pm$ 1.4 & 161 $\pm$ 2 & $-$ & 474.5 $\pm$ 29.9 & \sci{1.95}{-2}\\
\enddata
\end{deluxetable*}

\section{Analysis}\label{sec:discussion}
\subsection{Photometry and in-band spectral indices}
We calculated the photometry of the two CS disks and the CB ring for both weighting schemes.
In the case of the CS disks, the photometry was done by integrating all emission inside of the fitted single component 2D Gaussians.
In the case of the CB ring, we used the residuals, shown in \autoref{fig:map} (e, f), and an elliptical annulus aperture with inner radius of \mbox{46 au} and outer radius of \mbox{285 au} with an inclination angle of 45\degr{} and a position angle of 161\degr.
The calculated fluxes are presented in \autoref{tab:gauss}.
The total flux of the two CS disks and the CB ring is similar to the 1.3 mm SMA fluxes presented by \citet{Liu2018A&A...612A..54L}, which did not resolve the individual components.

We used the visibilities of our two continuum spectral windows to estimate the spectral index $\alpha$ using
\begin{equation}
\alpha = \frac{\ln(F_{\nu_1}) - \ln(F_{\nu_2})}{\ln(\nu_1) - \ln(\nu_2)},
\end{equation}
where $\nu_1 = 232.2$, \mbox{$\nu_2 = 216.9$ GHz}, and $F_{\nu}$ is the amplitude for each spectral window.
We found that the longer baselines have $\alpha\approx2$ while the shorter baselines have $\alpha$ values between 2 and 3.
The smaller value of $\alpha$ for the shorter spatial scales can be an indication of the presence of larger dust grains in the CS disks or that the system becomes optically thick at these short spatial scales.
We note that because the continuum spectral window frequencies are too close together there are large uncertainties in the estimation of the spectral index.
However, the high SNR of our observations provides enough to perform a comparative analysis between the short and long baselines.

For completeness, we calculated the $\alpha$ index using the 7 mm ($\sim$43 GHz) VLA observations from \citet{Lim2016ApJ...826..153L} and our uniformly weighted integrated fluxes.
We obtained values of 2.32 $\pm$ 0.03 and 2.26 $\pm$ 0.04 for the North and South disks, respectively.

\subsection{Mass estimates for the components}
The disks around FUor stars are more massive than the disks around T Tauri stars \citep{Cieza2018MNRAS.474.4347C}.
Because the previously mentioned $\alpha$ index values indicate that our disks are probably optically thick, our mass estimates must be interpreted as conservative lower limits.
We calculated the dust masses using the following formula:

\begin{equation}
M_{\rm dust} = \frac{f\,d^2}{\kappa\,B_\nu(T_{\rm dust})},
\end{equation}

where $f$ is the integrated flux, $d$ is the distance \mbox{(147 $\pm$ 5 pc)}, $\kappa$ is the dust opacity (\mbox{0.22 m$^2$ kg$^{-1}$}) and $T_{\rm dust}$ is the dust temperature.
We calculated the brightness temperature for the peak of each CS disk and used it as an approximation of the dust temperatures for the CS disks: \mbox{$T_{\rm dust, N}$ = 160 K} and \mbox{$T_{\rm dust, S}$ = 100 K}.
For the CB ring we used \mbox{$T_{\rm dust, R}$ = 80 K}, which is the dust temperature at the characteristic radius of our radiative transfer model (see below).
Assuming a 100:1 gas-to-dust mass ratio, we calculated the total disk mass for each component and the results are presented in \autoref{tab:gauss}.
These mass lower limits could increase by a factor of up to 3 due to beam dilution effects.
In the following section, we present a comparison our mass lower limits with the mass estimates from other FUor disks.

\subsection{Structure of the circumbinary ring}
In panels \emph{e} and \emph{f} of \autoref{fig:map} we show the residuals after the removal of the CS disk contributions for the uniform and natural weighting schemes, respectively.
As mentioned earlier, the brighter areas in the northern and southern parts of the circumbinary emission are due to geometrical effects of an inclined optically thin dust ring.

We tested this scenario by building radiative transfer models using \mbox{RADMC-3D}\footnote{\url{http://www.ita.uni-heidelberg.de/~dullemond/software/radmc-3d/}} \citep{Dullemond2012ascl.soft02015D} and comparing them with the natural weighted residuals.
Each model was constructed using the parametrization presented in \cite{Andrews2009ApJ...700.1502A}, in which the surface density was described as

\begin{equation}\label{eq:sigmar}
    \Sigma(R) = \Sigma_c \left(\frac{R}{R_c}\right)^{-\gamma} \exp\left[-\left(\frac{R}{R_c}\right)^{2-\gamma}\right],
\end{equation}

where $R_c$ is the characteristic radius, $\gamma$ is the surface density gradient and $\Sigma_c$ is the surface density at the characteristic radius which is given by

\begin{equation}
    \Sigma_c = (2 - \gamma) \frac{M_{\rm dust}}{2\pi R_c^2}.
\end{equation}

Finally, the scale height is described as $H_R = R h$, where $h$ is the normalized scale height defined as

\begin{equation}
    h = h_c \left(\frac{R}{R_c}\right)^\psi ,
\end{equation}

where $h_c$ is the scale height at $R_c$ and $\psi$ indicates the disk flaring.
The models were generated using two heating sources with \mbox{T$_*$ = 10,000 K} and \mbox{R$_*$ = 2 R$_\odot$}; these values are roughly equivalent to the total bolometric luminosity of the system and are representative of the enhanced accretion rate.
We fixed the positions of the two heating sources to the peak emission of each CS disk, and assumed the heating sources and the circumbinary ring to be in the same plane.
We constructed images based on each model using the best-fitted inclination and position angles presented in \autoref{tab:gauss}.

We searched for the best fitting model by running a Markov chain Monte Carlo (MCMC) using the \textsc{EMCEE} Python module \citep{ForemanMackey2013PASP..125..306F} to explore the parameter space and maximize the likelihood function.

\begin{figure*}[ht]
\plotone{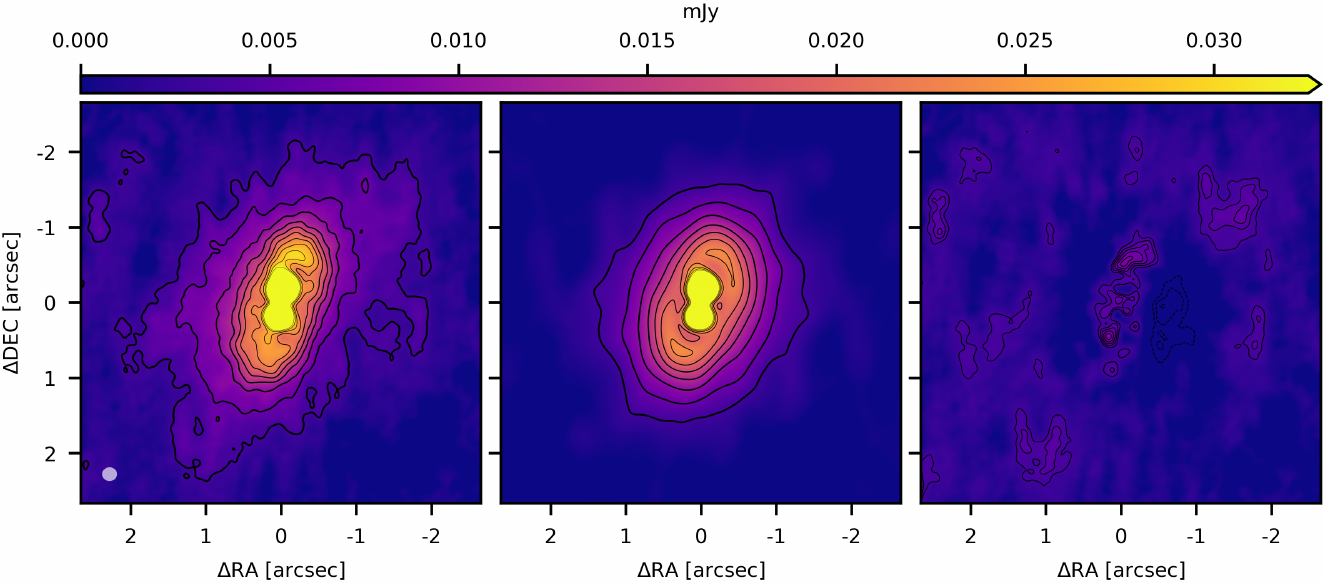}
\caption{Radiative transfer modeling of the CB ring based on the natural weighted image of L1551 IRS 5. The colorscale is saturated at 30$\sigma$ to show the structured of the circumbinary material. Left panel shows our observations. Middle panel shows the best fit model convolved with the clean beam. Right panel shows the differences between the observations and the best fit model. The overplotted contours are from 3 to 27$\sigma$ in steps of 2$\sigma$.
\label{fig:rt}}
\end{figure*}

In \autoref{fig:rt} we present the natural weighted image, the convolved best-fitted model and the respective residuals.
The residuals show emission in the inner parts of the CB ring, which can be explained by dust in the inner cavity, and explain why our fitted characteristic ring radius of $R_c = 110.9^{+0.12}_{-0.18}\,$au is larger than the peak in the radial profile found in \autoref{fig:deprojected_rp}.
Additionally, there is positive residual emission north and south of the CS disks and negative residuals to the east of the position of the CB ring; these can be explained by a non-axysimmetric dust distribution and will require more detailed modeling.
Finally, there are two areas with a $\leq4\sigma$ significance in the northeast and southwest of the residual map, which are attributed to emission from the envelope.

\section{Discussion}
L1551 IRS 5 is a FUor-like system with two CS disks of the same age and, therefore, presents an interesting opportunity to compare the two disks.
The northern disk is $\sim$2 times brighter and only slightly larger than the southern disk, therefore we interpret this as an indicative that the northern source is the eruptive star.
This is supported by L-band observations presented by \citet{Duchene2007}, which showed the northern disk as a point-like source with a surrounding nebulosity and did not detect the southern source.

We put the CS disks of L1551 IRS 5 into context by comparing their sizes and masses to the sample presented in \citet{Cieza2018MNRAS.474.4347C}, where they analyzed eight young eruptive stars in contrast to a sample of T Tauri stars previously presented by \citet{Andrews2010ApJ...723.1241A}.
Our measured radii, presented in \autoref{tab:gauss}, indicate that the L1551 IRS 5 CS disks are smaller than the resolved sources in the \citet{Cieza2018MNRAS.474.4347C} sample.
The lower limits for the disk masses of each CS disk fall within the range of expected masses of T Tauri stars with small radii \citep[Figure 6 of][]{Cieza2018MNRAS.474.4347C}.
However, because the CS disks in L1551 IRS 5 could be optically thick, it is possible that we are underestimating their total masses by up to 90\% \citep{Liu2018A&A...612A..54L}, which would place the L1551 IRS 5 CS disks masses with what is expected from other FUor-like systems.

\citet{Tobin2018ApJ...867...43T} analyzed a sample of 17 Class 0 and Class I multiple-star systems with separations of less than 600 au using ALMA 1.3 mm and VLA 9.1 mm data.
They did not detect circum-multiple dust emission around any of the Class I systems; however, closer binaries are more likely to have circumbinary material \citep{Bate2000MNRAS.314...33B} and the closest separation for Class I systems in their sample is $\sim$90 au.
The in-band $\alpha$ spectral index values for the L1551 IRS 5 CS disks are similar to the spectral indexes presented by \citet{Tobin2018ApJ...867...43T}.

L1551 NE is a Class I protostellar binary system with a similar morphology to L1551 IRS 5: two CS disks, with a 70 au separation, and a CB ring.
Based on the ALMA 0.9 mm continuum images from \citet{Takakuwa2017ApJ...837...86T}, the biggest difference between the two systems is in their circumbinary ring brightness distribution.
Contrary to the relative axisymmetry of the L1551 IRS 5 CB ring, the western side of L1551 NE is much brighter than its eastern side, which only shows a few small clumps of material.
Additionally, the CS disks in L1551 NE are misaligned and not coplanar with the CB ring, which the authors suggested as an indication that the outer material from the CB ring is falling directly into the surfaces of the CS disks, instead of smoothly connecting with the outer parts of the disks.

\section{Summary}
In this work we have presented ALMA 1.3 mm continuum observations of L1551 IRS 5, a FUor-like object made up of two low-mass stars, each with a circumstellar disk, and a circumbinary ring.
The high-resolution of our observations has enabled us to clearly separate the three components of this system.

Having two circumstellar disks in the same FUor-like system opens up the possibility of analyzing each disk separately.
From the marginally resolved detection of the CS disks, we measured their radii and found that both of them are smaller than any of the previously resolved FUors at this wavelength.
The spectral index $\alpha$ indicated that the CS disks may be optically thick.
We calculated conservative lower limits for the total disk masses and found that they are in accordance with the expected values of other known FUor disks.

Based on the residuals, the circumbinary emission appears to be ring-shaped with a width of 38 $\pm$ 0.6 au.
From our radiative transfer modeling, the CB ring appears to be mostly axisymmetric with a smooth surface density.
We interpreted the more significant residuals as indication of small asymmetries in the circumbinary dust distribution.

Due to the complexity of the system and its distance, complementary observations with higher angular resolution and at different frequencies are necessary to 1) fully analyze the circumstellar disks, including their optically thick inner parts, 2) verify the presence of the third bright component and study its nature, 3) study the asymmetries found in the circumbinary material.

\acknowledgments
This project has received funding from the European Research Council (ERC) under the European Union’s Horizon 2020 research and innovation programme under grant agreement No 716155 (SACCRED).
This paper makes use of the following ALMA data: ADS/JAO.ALMA\#2016.1.00209.S.
ALMA is a partnership of ESO (representing its member states), NSF (USA) and NINS (Japan), together with NRC (Canada) and NSC and ASIAA (Taiwan) and KASI (Republic of Korea), in cooperation with the Republic of Chile.
The Joint ALMA Observatory is operated by ESO, AUI/NRAO and NAOJ.
On behalf of the SACCRED project we thank for the usage of MTA Cloud (\url{https://cloud.mta.hu/}) that significantly helped us achieving the results published in this paper.

\vspace{5mm}
\facilities{ALMA}

\software{astropy \citep{astropy}, CASA \citep{McMullin2007ASPC..376..127M}, EMCEE \citep{ForemanMackey2013PASP..125..306F}, RADMC-3D \citep{Dullemond2012ascl.soft02015D}}

\end{document}